# Capacitive Mechanism of Oxygen Functional Groups on Carbon Surface in Supercapacitors

Yi-Tao He*

*Room 4-304, No.5 Meishan Road, Hefei City, Anhui Province, P. R. China.*

*Corresponding author: electro_yitao@163.com

**Abstract**

Oxygen functional groups are one of the most important subjects in the study of electrochemical properties of carbon materials which can change the wettability, conductivity and pore size distributions of carbon materials, and can occur redox reactions. In the electrode materials of carbon-based supercapacitors, the oxygen functional groups have widely been used to improve the capacitive performance. In this paper, we not only analyzed the reasons for the increase of the capacity that promoted by oxygen functional groups in the charge/discharge cycling tests, but also analyzed the mechanism how the pseudocapacitance was provided by the oxygen functional groups in the acid/alkaline aqueous electrolyte. Moreover, we also discussed the effect of the oxygen functional groups in electrochemical impedance spectroscopy.

Keywords: oxygen functional groups; supercapacitor; carbon

# 1. Introduction

Due to the high energy density and long service life, electrochemical supercapacitors (ES) have got many attentions from researchers[1]. And as a kind of energy storage devices, the ES can be used for portable electronic equipments, low emission hybrid electric vehicles, and urban green bus, etc.[2].

Porous carbon materials have the advantages of large specific surface area, high conductivity, stable structure and relatively low price[3], making it an ideal electrode material for supercapacitors.[4, 5] According to the formula $C = S\varepsilon/d$, the capacitance $C$ is proportional to the surface area of the material $S$ and the solution dielectric constant $\varepsilon$, and is inversely proportional to the thickness of the double layer $d$.[6] And it is generally believed that the thickness of the electric double layer formed on the electrode surface in aqueous solution is only about one water molecule thick.[7, 8] Therefore, the larger specific surface area and the smaller thickness of the double layer of the carbon materials make it have more capacity than the traditional physical capacitors.[9]

# 2. An Overview of the Contribution of Oxygen Functional Groups to Capacitance

In the study of the capacitive characteristics of carbon materials, researchers have realized that the oxygen functional groups have a great influence on the electrochemical properties of carbon materials.[10, 11] But the study of the specific contribution of oxygen functional groups to it has gone through a long process. Hsieh and Teng[12] studied the problem systematically in 2002. They utilized a combination of temperature programmed desorption (TPD) method for studying the oxygen functional groups and electrochemical characterization, and obtained the changes in capacitance of carbon fibers in acidic aqueous electrolytes when the gas content of CO and $CO_2$ were in different percentages. Combined with earlier studies of TPD, it can be learned that CO is mainly derived from hydroxyl, carbonyl and quinone groups, while $CO_2$ is mainly comes from carboxyl, anhydride and lactone groups.[13] It can be drawn from the results, the groups converting into CO offered larger capacitance than that converting into $CO_2$, and the relationship between the content of CO or $CO_2$ and the specific capacitances was:[12]

CO content → capacitance: 0.58 mmol $g^{-1}$ → 7.032 F $g^{-1}$   (1)

$CO_2$ content → capacitance: 0.08 mmol $g^{-1}$ → 0.968 F $g^{-1}$   (2)

Based on this result, it can be calculated that the specific capacitance provided by the oxygen functional groups that converted into these two gases were:

CO: 12.12 F $mmol^{-1}$ (3)

$CO_2$: 12.10 F $mmol^{-1}$   (4)

Therefore, in acidic aqueous electrolytes, the capacitance provided by the hydroxyl group on the surface of carbon fibers was larger than that provided by the carboxyl group. Moreover, the capacity of each oxygen functional group could also be calculated according to experimental data, and the result was approximately $2.00925209 \times 10^{-26}$ F. With the success of Hsieh, many related researchers have begun to use the TPD technology to study the capacitive contribution of oxygen functional groups that on the surface of carbon materials.[14, 15]

In order to explain the reasons for the contribution of oxygen functional groups to the capacitance, in 2003, Nian and Teng[16] have analyzed the effect of the number of oxygen functional groups on the properties of the electrode by analysing of the Nyquist diagrams that obtained by electrochemical impedance spectroscopy (EIS).

The authors pointed out that the Nyquist diagrams of the electrode materials in double layer capacitors were mainly composed of two parts. One was a semi-circle in high frequency regions, which represented the contact resistance between the carbon materials and the current collector.

It is worth our attention that, in the study of pseudocapacitive materials or other batteries, this semi-circle represents the charge transfer impedance at the electrode surface/electrolyte interface.[17] When studying the EDL electrode material, however, assuming that there is no electron transfer in the electrochemical double layer, the semi-circle represents the charge transfer impedance at the interface of the current collector/electrode material, which is referred to herein as the contact resistance. Therefore, according to the spectrum, Nian found that the contact resistance raised with the increase of the oxygen functional groups, leading to the reduce of the conductivity of the carbon surface, which was one of the negative factors of the oxygen functional groups.

The another part of the Nyquist diagrams was a straight line in low frequency regions. When the line is more vertical, indicating that the electrode material have capacitive characteristics. That was because when the frequency approached zero ( $\omega \to 0$ ), $Z(0) = 1/j\omega C$ ,[18] and in this circumstances, the impedance was only related to the imaginary part, and had nothing to do with the real part, so the image was a straight line.[19] However, results from Nian showed that the straight line in low frequency regions usually deviated from the vertical position. In addition, he proposed to use the constant phase element (CPE) to replace the capacitance element in the traditional RC circuit to study this phenomenon. Certain results have been obtained by using this kind of new RC equivalent circuit (see in Figure 1) to simulate carbon material electrodes that rich in oxygen functional groups.

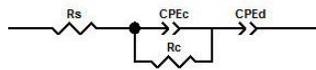

Figure 1. Equivalent circuit diagram that described the carbon material electrodes that riched in oxygen functional groups by Nian and Teng[16].

In this figure, as described by the authors, $R_s$ and $CPE_d$ represent solution resistance and double-layer impedance, and the paralleled $CPE_c$ and $R_c$ represents the current collector/carbon interface. The obtained Bode diagram showed that time constant values of the RC circuit increased with the increase of the number of oxygen functional groups, making a delayed response of carbon materials to ac signals. Moreover, the lower the frequency of the ac signal, the higher the capacitance measured by the Nyquist diagram of the carbon materials. As said before, the real part of the vertical line of the ideal capacitance was not related to the frequency. Therefore, the variation of capacitance with frequency manifests itself through the deviation from the vertically oriented line. Considering the capacitance of carbon material was closely related to the specific surface area, especially the pore volume,

Nian attributed this phenomenon to that the oxygen functional groups reduced the effective pore volume of microporous carbon materials, that was, the oxygen functional groups would hinder the migration of ions into micropores, which was another disadvantage of the oxygen functional groups.

The same year, Lozano-Castello et al.[20] indirectly verified Nian's point of view when studying on the characteristics of electrochemical double layer capacitance (EDLC) of carbon materials in non-aqueous electrolytes. The experimental results showed that the micropores on the surface of carbon materials were an important area for the formation of double layers, but the specific surface area of carbon materials was not linearly with the capacitance values, showing that there were other important factors affecting the capacitance values. Therefore, the author also studied the relationships between the number of oxygen functional groups, the micropore volume and the capacitance value by using the TPD technology, and found that the presence of oxygen functional groups would reduce the micropore volume of carbon materials. Moreover, the results by Lozano-Castello also showed that the influence of oxygen functional groups was greater than that of the specific surface area to carbon materials. This conclusion had also been confirmed by Yamashita et al.'s[21] experiment. As for why the oxygen functional groups would increase the capacitance value, the three authors all believed that the improved wettability of the carbon materials by oxygen functional groups made the ions in electrolyte easier to form electrochemical double layer. It can be seen that these works laid a good foundation for the study of the influence of oxygen functional groups, however, the research on electrochemical impedance was still not deep enough.

P. Simon's group[22] further deepen the research on the impedance. In 2004, they fabricated the electrodes for supercapacitors by using the modified Al collector and carbon materials, and they improved the equivalent circuit diagram simultaneously. Simon's point of view was similar to Nian's. They both considered that the semicircle in the high frequency region (parallel RC circuit of resistance and capacitance) represented the interface between carbon materials and the current collector. On the basis of Figure 1, a frequency-dependent resistance $R(\omega)$ was connected in series with the circuit, which was a good complement to Nian's circuit diagram. Moreover, the double layer capacitance C(ω) (CPE$_d$ as mentioned earlier) also changed with frequency (Figure 2).[23] Thus, the circuit diagram was closer to the fact that the low-frequency linear had deviated from the vertical position which was found in actual tests.

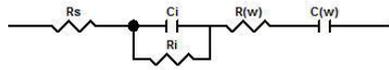

Figure 2. The improved equivalent circuit that describing the carbon material electrode with oxygen functional groups by P. Simon[22].

In addition, Simon also proposed that the Nyquist diagram of the carbon materials could be divided into two segments. As the same with what Nian had described that the Nyquist diagram contained a semicircle in high frequency regions and a straight line in low frequency regions. However, in Simon's conclusion, a line segment with 45° degrees angle appeared at the mid-frequency (the junction between the straight line and the semicircle). And he explained it as the occurrence of the mid-frequency region was related to the ions approaching to the surface of the carbon materials, which mainly referred to de Levie's research results.[24, 25]

After the results of EIS spectrum analysis and equivalent circuit diagram were basically determined, in 2005, Okajima's research group[26] further studied on the contribution of oxygen functional groups to the capacitance. They introduced oxygen functional groups on the surface of carbon materials by using plasma, and then studied the relationship between the oxygen functional groups and the capacitance value using Boehm's method.

Results showed that the capacitance value was proportional to the number of quinone groups within the range of oxygen supply in acidic aqueous electrolytes, illustrating that quinone was a key factor affecting the capacitance of carbon materials. As the quinone was one of the groups providing CO gas, so this conclusion was consistent with the previous results of Teng's work, which showing that the oxygen functional groups that provided CO gas could offer more capacitance. The reason for this phenomenon was that the quinone group was a kind of electron acceptor, and the increase in the capacitance might be due to the redox reaction of the quinone.

Was the oxygen atom in the oxygen functional groups really an electron acceptor? A similar discussion about nitrogen functional groups could be used as a reference. It is interesting to note that although there have been quite a number of papers that had provided evidences to prove that under the influence of electronegativity, the N atom in the nitrogen functional groups was an electron acceptor, which lead to the increase of the positive charge density of the neighboring C atoms.[27] However, in Hou et al.'s[28] point of view, the N atom contained one more electron than the C atom, which could enter the delocalized π-bonds which adjacent to the C atom. So the results showed that the total effects made the N atom actually an electron donor.[28] From these two diametrically opposed viewpoints, Okajima's conclusion seemed to be applicable only when the oxygen functional groups was not in the range of the delocalized π-bonds formed by C atoms. In addition, in order to measure oxygen content, X-ray photoelectron spectroscopy (XPS) test had been introduced by Okajima, which also made the measurement of oxygen content more precise. In 2006, Oda's team[29] accurately compared the contribution of phenolic hydroxyl groups and carboxyl groups to the capacitance by using the XPS test. The experimental results showed that the growth rate of the capacitance became more sensitive to the number of phenolic hydroxyl groups with the increase of the number of these two groups. As mentioned earlier, hydroxyl is also one of the groups that could provide CO gas. And enough evidences had shown that the capacitance provided by the hydroxyl groups and the quinone groups was more than that provoided by the carboxyl groups in acidic aqueous electrolytes.

In order to provide the mathematical relationship between the number of oxygen functional groups and the capacitance value, Centeno and Stoeckli[30, 31] had made a very valuable work. They further determined that the current was inversely proportional to the capacitance on the basis of the Qu and Shi's[32] work. That was due to the fact that the ions in the electrolyte were too late to enter into the pores of the surface of the carbon materials under large current, thus could not form enough electrochemical double layers. Combined with the two unfavorable factors of oxygen functional groups proposed by the Teng's group, Centeno and Stoeckli's experimental results showed the presence of oxygen functional groups was an important reason for blocking ions from entering the pores at large current. After summing up the results of many other researchers, they obtained the relationship between the total oxygen content of the carbon materials and the capacitance value:[31]

$$C\,(\mathrm{F\,g^{-1}}) = C_0 \exp[-5.32 \times 10^{-3} d(1+0.0158[O]^2)] \quad (5)$$

Where $C_0$ represents the coefficient, $d$ represents the current density, and [O] represents the total oxygen content. Although the formula can not apply to all types of oxygen functional groups, it gives a general relationships between oxygen content, current density and the capacitance. In other words, the formula showed that the more oxygen functional groups on the surface of carbon materials, the worse the rate performance of the electrode. As the rate performance of the electrode is closely related to the internal resistance of the electrode,[33] and from the above discussion, it is found that the oxygen functional groups could increase the impedance of the interface between carbon materials and the current collector, that is, the internal resistance of the electrode increased. So it is reasonable that the rate performance had been reduced.

## 3. Analysis of the Capacitance Growth of Carbon Materials With Oxygen Functional Groups in Cyclic Tests

Unfortunately, the above works had not been further studied on the problem of the oxygen functional groups hindering the ions from getting into the pores. And the problem was closely related to the variation of the capacitance value of the carbon materials electrode in cyclic tests. In order to get a further study on this, we had performed the cyclic charge/discharge tests with carbon materials and performed the EIS tests before and after the tests. The results were shown in Figure 3. It needs be noted that all the experimental results in this article came from author's previous

works.[34, 35]

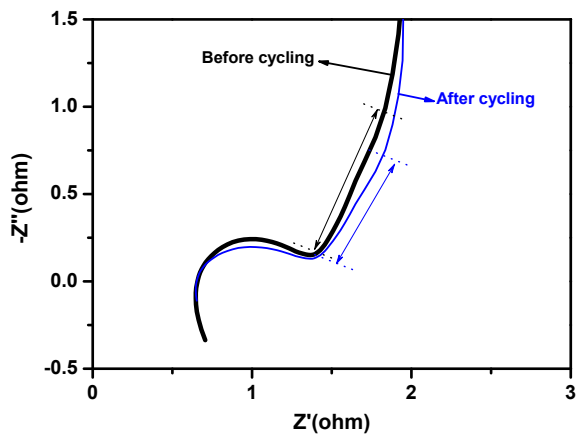

Figure 3. The Nyquist plots before and after 10000 cycles test of carbon material in 6 M KOH.

It can be seen from Figure 3 that long time charge/discharge process had a great influence on the electrode impedance spectrum. Especially that the slope of the line segment in the intermediate frequency region became smaller after the cycling. According to the conclusion of Simon, the line represented the impedance of ions entering into the surface of carbon materials in electrolyte. So, the smaller slope showed that cycling process made it easier for ions entering into the pores. Meanwhile, some papers had reported that the capacitance of carbon materials with oxygen functional groups would get a increase after long-term usage.[36] This showed that the factor that ions were more easily to enter the pore had greater impact on the capacitance value than that of the effect of the increasing electrode resistance (which was performance as the radius of the semicircle in the high frequency region increased after the circulation). As mentioned earlier, the wettability of carbon materials could be improved by the oxygen functional groups, which was helpful for enhancing the ion diffusion in aqueous electrolyte. And many researchers believed that this was the main reason for the increase in the capacitance of carbon materials in cycling process, which could be shown observably in the Nyquist diagram. In 2007, Fang[37] studied on the changes in the value of each component in Nyquist diagram and the equivalent circuit when the wettability of carbon materials was affected by the oxygen functional groups. Here, the equivalent circuit proposed by Simon was also applied. Fang found that the trend that the value of capacitance decreased with the change of frequency slowed down in the low frequency region and the straight line in the low frequency area in the Nyquist diagram became more vertical and approached to the ideal capacitance behavior when the surface wettability of carbon materials was enhanced. It showed that the wettability increased the migration rate of ions on the surface of carbon materials, and reduced the mass transfer resistance, making it easier for ions to form a double layer, showing that the improvement of wettability was indeed one of the reasons for the increase in capacitance.

While after testing the cyclic voltammetry (CV) curves of carbon electrode before and after the cycling tests (see Figure 4), an interesting phenomenon had been found.

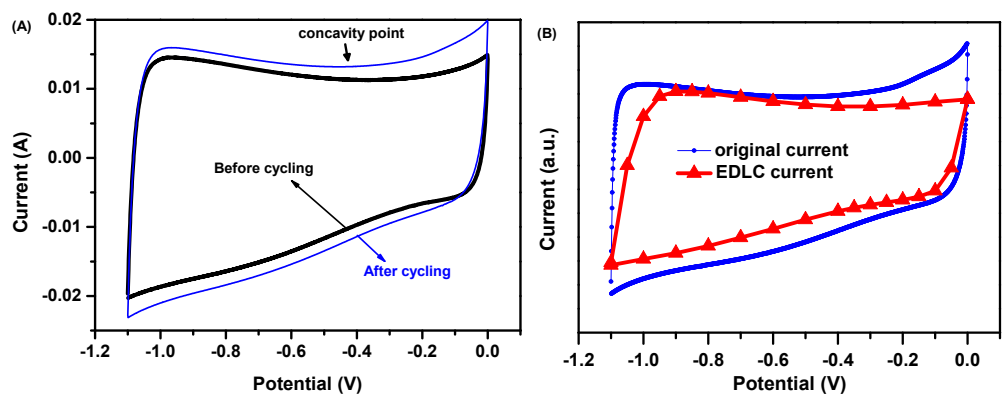

Figure 4. (A) The CV curves of carbon electrode materials before and after the cycling tests in KOH electrolyte; (B) Separation of double layer current and redox current in CV curves of carbon electrode materials.

The area of CV curves measured after the cycling was larger than that of the CV curves before the cycling. And the increased area was mainly concentrated in the high potential of the charge curves and the low potential of the discharge curves. According to our previous work of separating the double layer current and redox current from CV curves of carbon electrode materials,[38] the pure EDLC current could be separated from the original CV curves, as shown in Figure 4 (B). The separation results showed that the pseudocapacitance provided by the oxygen functional groups was also mainly concentrated in the high potential of the charge curves and the low potential of the discharge curves in alkaline aqueous electrolytes. So that the contribution of oxygen functional groups to the increased capacitance value could not be ignored in cycling process. It was not only easier for the electrochemical double layer to form, but the oxidation/reduction reaction was also more likely to occur for the oxygen functional groups when the wettability had been increased, showing that the mechanism for the oxygen functional groups providing the pseudocapacitance was more likely be involved by the ions in the electrolytes. It should be noted that, in 2000, Frackowiak[6] had noted that the pseudocapacitance area in the CV curves of the carbon materials had decreased after the cycling process, showing that the oxidation/reduction processes of the oxygen functional groups was a quasi-reversible process, and the capacity would gradually decrease as the process proceeded. In general, the capacitance of carbon materials with oxygen functional groups in cycling process would get a decrease at last rather than maintained a rise.

In addition, an important information from the CV curves in Figure 4 (A) could also be got. Research results from Y.-H. Wen et al.[39] showed that the CV curves of carbon materials would have a significant concave at the point zero charge. And the degree of the concave point was closely related to the utilization of micropores on the surface of the carbon materials. In different cases, the absolute value of the difference between the current value at the highest voltage point and the current value at the concave point can be used to evaluate the degree of fluctuation. Wen found that the degree of the concave point and the resistance of ions migrating to the pores decreased with the increase of the the pore size of carbon materials. While in Figure 4 (A), the degree of the concave point showed a slight rise after the cycling, showing that the pore utilization ratio of carbon materials had improved, making ions into the pore easier.

Therefore, the phenomenon of capacitance growth of carbon materials with oxygen functional groups should be attributed to three reasons: (*i*) the surface wettability of carbon materials had been improved by the oxygen functional groups; (*ii*) with the improvement of wettability in cycling process, it was easier for the oxygen functional groups to contact with the ions in the electrolyte, thus making the oxidation/reduction reaction more easily to occur; (*iii*) the pore utilization ratio of carbon materials would increase gradually in cycling process.

**4. Influence of Concentration Polarization**

The variation rules of CV curves of oxygen functional groups within the pH range of 0~14 in aqueous solution were studied by Andreas and Conway[40] in 2006. Andreas found that an obvious redox current peak near 0.55V appeared in the CV curves when the electrolyte was acidic, and the appearance of the peak had no relationship with the kind of the acid, but only related with the concentration of $H^+$. When the concentration of $H^+$ decreased gradually, and the pH value increased, the redox peak of oxygen functional groups disappeared gradually. Combined with previous works by other researchers,[41] Andreas got the conclusion that the concentration of $H^+$ was the key factor for the redox reaction of oxygen functional groups. The disappearance of the redox peak in alkaline solution was not due to the interaction of $OH^-$ with the oxygen functional groups, but because of the lack of $H^+$. Another important phenomenon was that the redox potentials of the oxygen functional groups shifted to the positive potential with the increase of the pH value.

According to the fundamental knowledge of electrochemistry, the carbon materials electrode was a part of the electrolytic cell in the testing process of CV curves. While in the electrolytic cell, the concentration polarization might have a great influence on the system in sometimes.[42] Referring to the results of Andreas:[40] When pH was 0, the redox potential was 0.5655 V (vs. RHE), and the peak current was 304.406 mA g$^{-1}$; When pH was 1.02, the redox peak potential was 0.6485 V (vs. RHE), and the peak current was 293.678 mA g$^{-1}$. The peak potential difference was 83 mV at different pH values. When pH was 0, assuming that the concentration of $H^+$ was large enough, thus the current density was the limiting current density. We can use the cathodic concentration polarization equation:

$$\Delta \varphi = \frac{RT}{zF} \ln(1 - \frac{i}{i_d}) \quad (9)$$

Here *φ*, *R*, *T*, *z*, *F*, *i* and *i$_d$* represented the gas constant, the temperature, the electron transfer number, the Faraday constant, the actual current density and the limiting current density, respectively.

As only one electron had been transferred in the oxidation-reduction reaction for each oxygen functional group, so here $z=1$, putting the data into the equation:

$$\Delta\varphi = \frac{8.314 \text{J}\cdot\text{mol}^{-1}\text{K}^{-1} \times 298.15\text{K}}{1 \times 96485 \text{C}\cdot\text{mol}^{-1}} \ln(1 - \frac{293.678 \text{mA}\cdot\text{g}^{-1}}{304.406 \text{mA}\cdot\text{g}^{-1}}) \quad (10)$$

The approximately calculated results were that $\Delta\varphi = -86$ mV. When the carbon materials were used as the negative electrode, the redox potential of the oxygen functional groups would get a shift of 86 mV to the positive direction, which was very close to the experimental results of Andreas's. Therefore, we believe that the concentration polarization was an important reason for the shift of the redox peak potential caused by pH value in Andreas's results.

## 5. Analysis of Capacitive Mechanism of Oxygen Functional Groups in Acid/Alkaline Electrolyte

From the above discussions, and many literature of oxygen-rich carbon-based electrodes, it is quite obvious that the electrochemical characteristics of the carbon material in the acidic aqueous electrolytes and alkaline aqueous electrolytes are not exactly the same. This is mainly manifested in the presence of redox peaks in acidic aqueous electrolytes, which can be clearly observed in the CV curves. While in alkaline aqueous electrolytes, there is no apparent redox peaks.[43] These facts show that the mechanism by which the oxygen functional groups provide pseudocapacitance are different in acidic/alkaline aqueous electrolytes, which need to be discussed separately.

### 5.1 In acidic aqueous electrolytes

*5.1.1 Pseudocapacitive mechanism of hydroxyl*

In the acidic aqueous electrolytes, a redox reaction have been proposed by scholars for pseudocapacitance provided by oxygen functional groups. Kinoshita[44] mentioned the following reaction in his works:

$$> C_xO + H^+ => C_xO//H^+ \quad (6)$$

Where the symbol > represents the surface of the carbon material, and the symbol // represents the adsorption interface between $H^+$ and the oxygen functional groups. When considering the electron transfer, Teng[12] referred to a following reaction formula:

$$> C_xO + H^+ + e^- => C_xOH \quad (7)$$

This reaction formula indicates that the proton $H^+$ is not only simply adsorbed on the oxygen functional groups, but involved in the transferring electrons. There are three possible cases: (*i*) the surface has no oxygen functional groups (OxyFG); (*ii*) the surface has OxyFG but no redox reaction; (*iii*) the surface has OxyFG occurring redox reaction. Then, on the basis of Eq.(7) and the ordinary double-layer model[6], we can draw the electrode surface model graph with OxyFG after formation of electrochemical double layer, as shown in Figure 5.

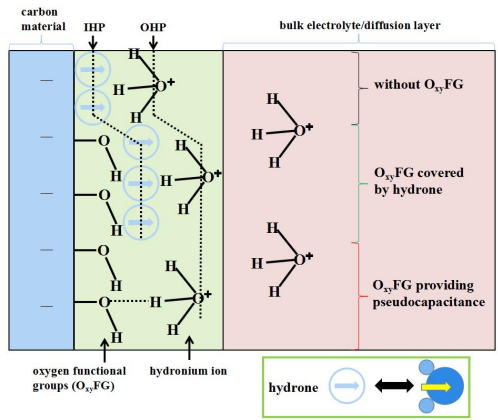

Figure 5. Electrochemical double layer model on the surface of carbon materials with oxygen functional groups in acidic aqueous electrolytes.

Take hydroxyl contributing to the most capacitance as an example. It is assumed that the carbon material, in $H_2SO_4$ aqueous solution, is the cathode and its surface is negatively charged. Hydronium ions ($H_3O^+$) are present in electrolyte around the carbon surface. In the places where OxyFG cannot occur redox reaction, the hydrones form the inner Helmholtz layer (IHP), and are capable of polarizing under electric fields. Meanwhile, $H_3O^+$ form outer Helmholtz layer on the whole surface of electrode. The preferential orientation of positive charge center in $H_3O^+$ is close to the negatively carbon surface. In the other places where OxyFG can occur redox reaction, The O atom on the hydroxyl group forms a bond with one of the H atoms in $H_3O^+$. We could use density functional theory (DFT) to calculate the most stable configuration of this macromolecule under the electric field. When constructing the model, we considered the situations that the carbon surface are formed by several benzene rings behind the hydroxyl group, and in water solutions.

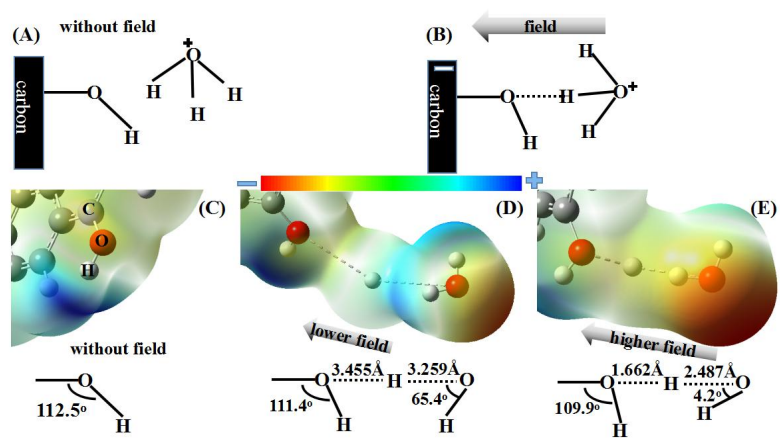

Figure 6. (A) There is no bonding between hydroxyl and $H_3O^+$ when there is no electric field applied; (B) Forming a bond between hydroxyl and $H_3O^+$ under the electric field; (C-E) Electron cloud distributions, the changes in bond length and bond angle in this macromolecule under the zero-lower-higher fields.

In Figure 6 (A), when there is no electric field applied, the results of calculating showed that hydroxyl does not have hydrogen bonding interaction with the group and the $H_3O^+$, and there is no stable molecular configuration with lowest energy. After applying the electric field, the surface of the carbon material is negatively charged, as shown in Figure 6 (B). The positive charge center consisting of three hydrogen atoms in $H_3O^+$ exhibit the closer approach to the carbon surface. The calculated results have strong regularities in changes of molecular configuration: (*i*) in the absence of external electric field (Figure 6 (C)), the H atom of the hydroxyl group is the positive charge center, furthermore, the electrons in the oxygen atom are involved in the construction of large π-extended conjugation on the electrode surface, which is similar to the study for N atom reported by Hou[28]; (*ii*) when a lower electric field (2.5 V/Å) was applied (Figure 6 (D)), the C-O-H bond angle in the hydroxyl group and the H*-O-H bond angle in $H_3O^+$ become simultaneously smaller, where H* represents the H atom involved in the hydrogen bond. (*iii*) when a higher electric field (5.1 V/Å) was applied (Figure 6 (E)), it is obvious that the degree of reduction of the H*-O-H bond angle is much larger than that of C-O-H. The H atom in H*-O-H can be almost considered as an absorbed atom on H*⋯O bond. In addition, the length of the O⋯H*⋯O bond linking hydroxyl and $H_3O^+$ become shorter under high electric field, which indicates that the enhanced field can make the interaction between the two molecules much stronger, and their electron clouds are approaching each other. The electrons in hydroxyl have a tendency to shift toward the oxygen atoms in $H_3O^+$, and a negative charge center appears on the oxygen atom. The end result is that an electron is lost from the hydroxyl according to the Eq.(7). We can use a reaction equation that can better express this process:

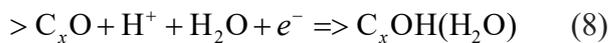

$$>C_xO + H^+ + H_2O + e^- => C_xOH(H_2O) \qquad (8)$$

*5.1.2 A brief introduction of quinone providing pseudocapacitance*

At the same time, there is also another view that the quinone is the group providing more pseudocapacitance than others. As mentioned above, The experiments conducted by Okajima's group[26] show that the increase in the capacitance caused by the presence of the oxygen functional groups is due to the redox reaction of the quinone. Andreas[40] also mentioned that the the redox of oxygen functional groups in acidic aqueous electrolytes is mainly the conversion between hydroquinone/quinone. In 2011, Roldán et al.[45] added hydroquinone into $H_2SO_4$ solution to prepare redox-active electrolyte by using the reaction, which greatly enhanced the capacitance value of carbon materials. In this case, the Eq.(8) is not appropriate for directly describing the quinone-based redox reaction. But Singh and Paul's reaction equation could provide a reference for this:[46]

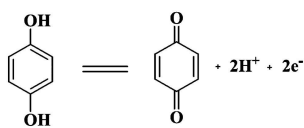

Scheme 1. Reaction equation of hydroquinone, described by Singh and Paul[46].

Moreover, DFT can also be used to make a simple optimization of the keto-$H_3O^+$ structure, as shown in the following figure:

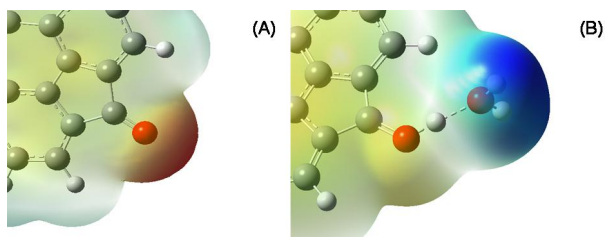

Figure 7. (A) There is no bonding between ketone and $H_3O^+$ when there is no electric field applied; (B) Forming a bond between ketone and $H_3O^+$ under the electric field.

It is calculated that, when the hydronium ion is bonded to the oxygen atom of the ketone group, the O⋯H* bond length is about 1.136 Å, which is very close to the O-H bond length (0.996 Å) of the hydroxyl group under the same situations. This also indirectly verifies the correctness of the reaction in Scheme 1. Hence, it can be concluded that the quinone group combines with H* in the $H_3O^+$ to transform the ketone group to the hydroxyl group, resulting in the movement of charge.

In general, in acidic aqueous electrolytes, the pseudocapacitance by quinone and hydroxyl is an most important contributor to the enhancement of a capacitance for carbon materials.

## 5.2 In alkaline aqueous electrolytes

Researchers have long been working on the capacitive properties of carbon materials in alkaline aqueous electrolyte. However, as the radius of hydrated $K^+$ is smaller[47] and it has a relatively large polarization intensity in aqueous solution, therefore, a larger charge density of the electrochemical double layer could be more conducive to the improvement of the capacitance when using KOH solution.

Although many researchers had reported that no obvious redox peaks were observed in the CV curves of carbon materials, by analysing the experimental phenomena, Cheng and Teng[48] thought that the redox reactions still existed in KOH solution. Cheng noticed that a wide reduction peak appeared when first tested the carbon materials with oxygen functional groups. And the reduction peak became smaller at the second scan and disappeared at the third scan. In addition, the size of the reduction peak was also changed with the number of oxygen functional groups when first tested the carbon materials.

In 2000, Barisci et al.[49] pointed out that the dissolved oxygen could affect the negative charge storage on the surface of carbon materials in alkaline electrolyte, thus shorten the discharge time. Considering the fact that the dissolved oxygen could affect the the capacitance characteristics of carbon materials in alkaline electrolyte, Cheng attributed the reduction peaks in the CV curves to the irreversible reactions of oxygen functional groups and speculated that the reaction process was: $O_2$ and $H_2O$ had reacted with the oxygen functional groups and formed $OH^-$ or $HO^{2-}$.

In fact, Yeager[50] had presented the electrochemical reduction reaction of dissolved oxygen or adsorbed oxygen on the surface of carbon materials in the field of electrocatalysis in 1984. In alkaline electrolyte, the reaction was:

$$O_2 + H_2O + 2e^- \rightarrow HO_2^- + OH^- \quad (11)$$

And this irreversible reaction was likely to be the cause of Cheng's experimental phenomenon. While as could be seen from many papers, no obvious redox peaks was existed in the CV curves of the carbon materials in KOH solution but only existed a current platform or concave points similar to the Figure 4 (A). In the case of Andreas's point of view that mentioned in the forth section, the surface of carbon materials were mainly depend on the electrochemical double layer to provide capacitance in KOH solution. As the radius of the hydrated $K^+$(3.31Å) was larger than that of the hydrated $H^+$(2.82Å), it would be affected by the oxygen functional groups more greater. Especially when in the micropores of carbon materials, it was more difficult for the hydrated $K^+$ to form an electrochemical double layer. Combined with the previous results achieved by Wen[39], it was also an important reason for the large fluctuations that occur in the current platform of the CV curves in carbon materials in KOH solution.

When the diffusion of ions in the electrolyte was affected by the pores and the oxygen functional groups, the interfacial tension between the carbon materials and the electrolyte then should be taken into account. As the differential capacitance curve was a useful tool for studying the electrochemical double layer on the solid/liquid interface,[51] which was used to describe the relationships between the quantity of electricity, the interfacial tension and the electrode potential. It could be used to describe the properties of the electrochemical double layer on the surface of carbon materials.

The differential capacitance curve was used to describe the relationship between $dQ/d\varphi$ and $\varphi$, and in the CV curves of carbon materials, $Q=\int idt$, so we can get $dQ/dt = i$. The electrode potential was $\varphi = \varphi_0 + vt$ when at a certain sweep rate of $v$, thus $d\varphi/dt = v$, so we can get $dQ/d\varphi = i/v$. It can be seen that the relation curves between $dQ/d\varphi$ and $\varphi$ was equivalent to the relation curves between $i/v$ and $\varphi$ (the current value of CV curves divided by the sweep speed). Thus in other words, the CV curves were thought to be another form of the differential capacitance curves.

In 2000, the CV curve was used as the differential capacitance curve to describe the relationship between the ion diffusion and the pore size on the surface of carbon materials by Salitra et al.[52] He also pointed out that the electrode potential at the lowest point on the CV curve (the concave point in the current platform) was the zero charge potential (zcp). In the testing process of CV curves, the amount of the hydrated $K^+$ in the electrochemical double layer on the surface of carbon materials increased with the increase of the electrode potential. The surface area of the hydrated $K^+$ layer would be minimized under the influence of electrostatic repulsion, and this process had reduced the influence of the interfacial tension. Therefore, the interfacial tension would be changed with the change of the electrode potential. However, the interfacial tension would hardly be changed with the change of the electrode potential at the zcp, making it the lowest degree of charge separation on the electrode surface, and thus the hydrated $K^+$ diffused into the bulk solution. Therefore, the capacitance was the smallest at this moment, and thus made an S-shaped CV curve of carbon materials.

The increase of oxygen functional groups could increase the amplitude fluctuations of CV curves, but there was no obvious changes in the position of zcp, as shown in Figure 8.

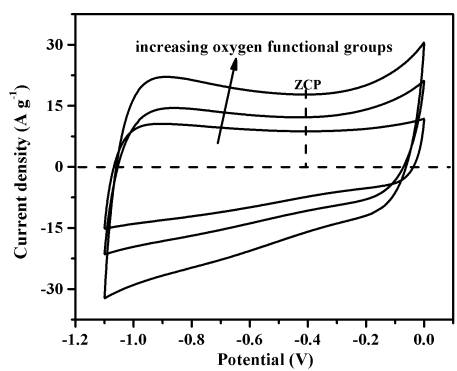

Figure 8. The amplitude fluctuations of CV curves increased with the increase of the number of oxygen functional groups (in KOH electrolyte).

The oxygen functional groups had a great influence on the migration of ions and prevented them from entering into the pore channel, which enhanced the diffusion of ions into bulk solution in the electrochemical double layer. This might be the reason why the CV curves of carbon materials presented an S-shaped curve, and the conclusion was consistent with that of Wen's which mentioned in the second section.

In addition, according to the results of the separation of the electrochemical double layer current and the pseudocapacitance current in Figure 4 (B), we could infer that: the pseudocapacitance increased after several times of charge/discharge cycles compared with the pure electrochemical double layer current in KOH solution. While the hydrated $K^+$ would be more easier to enter into the pore in cycling process, which might be attributed to the increased capacitance in the high potential of the charge curves. Therefore, the generation of pseudocapacitance might due to the insertion/deinsertion reaction of ions in the pore (increased the pore utilization ratio), and improved wettability.

Although the radius of the hydrated $K^+$ was large, Chmiola et al.[53] argued that the hydrated cations appeared to be squeezed, and the twisted hydrated cations would gradually enter into the interior of the hole to form an electrochemical double layer. Moreover, Huang[54] had constructed the model of the electrochemical double layer in micropores (apertures below 0.3 nm), which showed that the inference was reasonable.

Therefore, in KOH electrolyte, the CV curves of the carbon materials with oxygen functional groups could be divided into three segments, and each of which was correspond to the formation process of different electrochemical double layer, as shown in Figure 9.

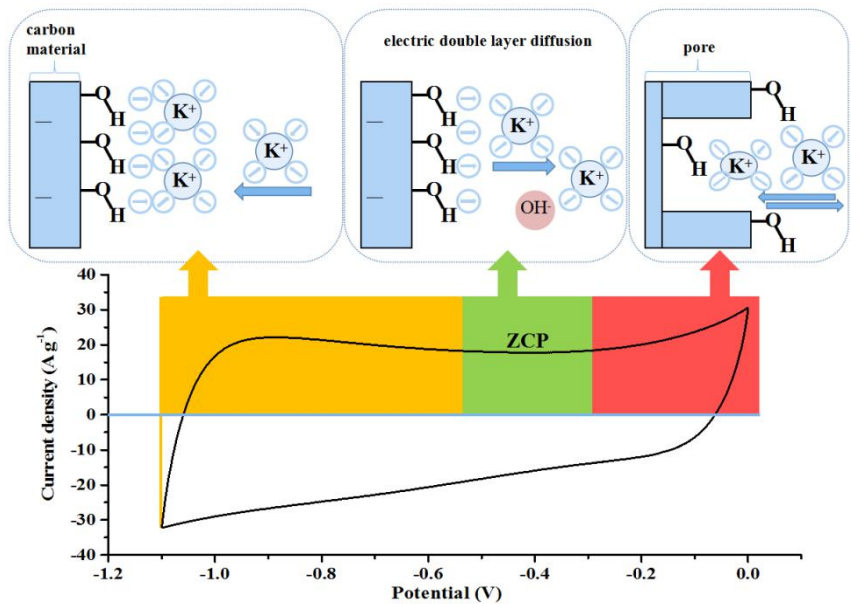

Figure 9. Schematic diagram of the formation of electrochemiacl double layer of CV curves of carbon materials with oxygen functional groups at different stages in KOH electrolyte.

As shown in Figure 9, when the charge current of the CV curves was in the low-potential region, the hydrated $K^+$ in KOH electrolyte would move rapidly to the surface of the electrode to form the outer Helmholtz layer (OHP). While in the middle-potential region, the interfacial tension reached maximum values, the hydrated $K^+$ in KOH electrolyte had been dissipated and the steric hindrance effect of the oxygen functional groups enhanced the diffusion of ions into the bulk solution, leading to a minimum capacitance. In the high-potential region, the hydrated $K^+$ appeared to be squeezed, and a quasi-reversible process of the insertion/deinsertion reaction occurred in the pore, which had generated an additional capacitance.

We could use the following equation to represent the quasi-reversible process:

$$>C_xO + [K(H_2O)_n]^+ + e^- \xrightleftharpoons{charge/discharge} >C_xOK(H_2O)_{n-y} \quad (12)$$

Where $n$ represents the number of water molecules that coordinated with $K^+$, and $y$ represents the number of water molecules that the hydrated $K^+$ had lost when been squeezed. As the behavior of oxygen functional groups in neutral solutions was between that in acidic solutions and in alkaline solutions and the capacitance was small, thus it would not be discussed in this article.

## 6. Effect of Oxygen Functional Groups on EIS

### 6.1 Determination of equivalent circuit

As mentioned above, the Nyquist plot of a porous carbon electrode could be divided into three frequency regions. Kötz[55] described in detail the connotation of each region, and he reached similar conclusions of Nian's[16] and Simon's[22]. Kötz also mentioned that the real-axis intersection of the impedance spectra represented the total effective series resistance (ESR). ESR of a capacitor was the internal resistance that appears in series with the capacitance of the device. The value of ESR is related to the resistance between leads and electrode or counter electrode, and the welding point, etc. The ESR is inevitably present in realistic capacitors. However, in order to eliminate interference from ESR it is necessary to use a same counter electrode, collector and same leads, which makes the ESR of work electrodes simply relevant to itself.

Although the equivalent circuit in Figure 2 had been improved, the Warburg impedance represented by a 45° line in the mid-frequency region was not taken into account. The reason why a typical 45° line in impedance plot appears is because, in deriving the diffusion impedance equation of a semi-infinite electrode, the product term (1-j) appears so that the real part is equal to the imaginary part.

Hence, if the diffusion of ions in the pores was taken into account, the Warburg impedance element must be introduced, similar to the changes made by Liu[56]. We compared the fitting results without or with Warburg impedance element, as shown in Figure 10.

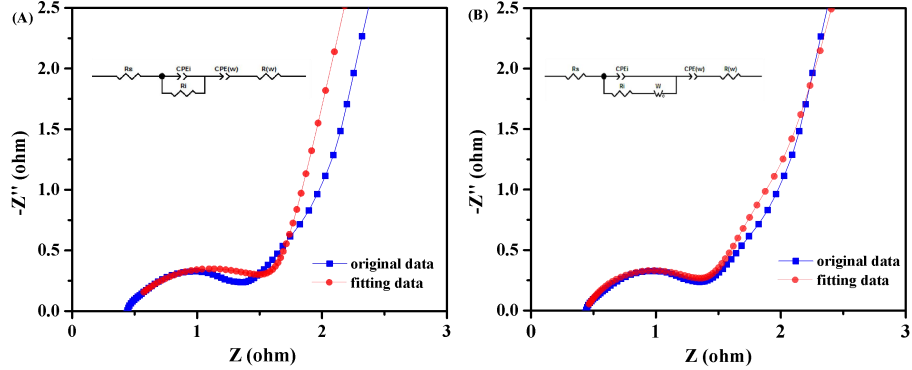

Figure 10. The fitting results without (A) or with (B) Warburg impedance element.

It can be clearly seen that the circuit in Figure 10 (A) does not fit the 45° line in the mid-frequency region. But the fitting result from the circuit in Figure 10 (B) is much better. In fact, the product term (1-j) in the diffusion impedance equation already indicates the necessity of the presence of the Warburg impedance element. The significance of the circuit elements in the actual electrode material could be illustrated by Figure 11.

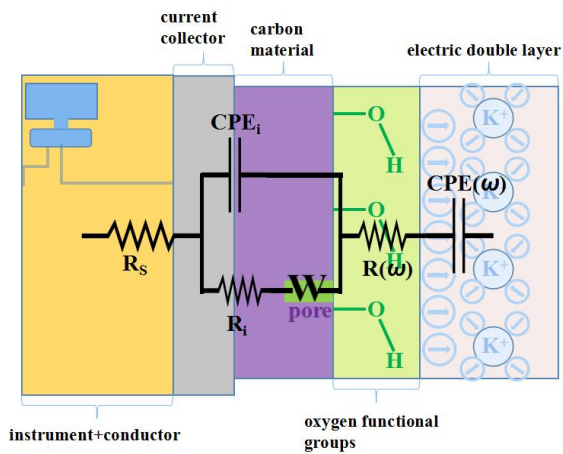

Figure 11. Equivalent circuit elements in realistic electrode system.

In Figure 11, $R_s$ equals to ESR. The basic parallel RC circuit, consisting of $CPE_i$ and $R_i$, represents the interface of the collector/carbon material, which is expressed as a semicircle at the high-frequency in the Nyquist diagram. This semicircle represents the electron transport at the interface of the collector/carbon material during the charging/discharging process of the electrochemical double layer. Moreover, $R_s$, $CPE_i$ and $R_i$ are constant under certain conditions, hence these values need to be fixed when fitting the Nyquist diagram in full frequency range.

A straight line with a slope of 45° from the real axis corresponds to semi-infinite Warburg impedance. The Warburg element is used to represent linear diffusion under semi-infinite conditions, which measures the degree of difficulty of entering the pores on carbon surface. $R(\omega)$ is a resistance that changes with frequency, which may be interpreted as resulting from surface porosity, solution conductivity and steric hindrance by the oxygen functional groups, etc. And $CPE(\omega)$ is related to the double layer capacitance on porous carbon surface, considering interfacial diffusion effect.

### 6.2 Influence of oxygen functional groups on Nyquist diagram

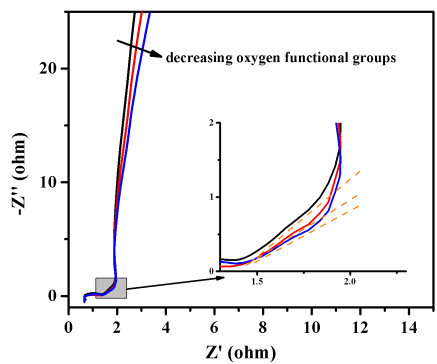

Figure 12. The permeability changes with the amounts of oxygen functional groups in Nyquist diagram.

As shown in Figure 12, the slope of the Warburg impedance tail is seen to increase with increasing the amounts of oxygen functional groups, which indicates an increase of ionic diffusion impedance in the pores. At the same time, the straight line at the low frequency gradually becomes more vertical. On the one hand, as previously described, too many oxygen functional groups will, to some extent, hinder the transport of ions into the pores, which increases the Warburg impedance. On the other hand, the strongly polarized oxygen functional groups also enhances the hydrophilicity of the surface of the carbon material, which is very beneficial for the rapid accumulation of ions on the electrode surface, but also reduces the mass transfer resistance. These performances are consistent with the content of the full text.

**7. Conclusion**

According to the foregoing analysis, the oxygen functional groups have a great influence on the capacitive performance of carbon materials. There are two main disadvantages: (*i*) decreased the surface conductivity of carbon materials; (*ii*) prevented ions from entering into the pore channel. And the main reasons for the increase of the capacity caused by the oxygen functional groups were: (*i*) the wettability of the electrolyte on the surface of the carbon materials had been enhanced; (*ii*) the Faraday pseudocapacitance which contributed by the electron transfer (redox reaction) could be provided; (*iii*) the pore utilization ratio of carbon materials would increase, e.g., the quasi-reversible process of the hydrated $K^+$ insertion/deinsertion reaction occurred in the pore.

In the acidic aqueous electrolyte, the electrons on the oxygen functional groups were attracted by $H_3O^+$, which lead to the separation of the positive and negative charges, thus making the redox reaction occurred. While in the alkaline aqueous electrolyte, the pseudocapacitance might be caused by the insertion/deinsertion reaction of the hydrated ions in the pore. Finally, the influence of the number of oxygen functional groups on the Nyquist diagram had also been discussed. The experimental results showed that the mass transfer resistance decreased with the increase of the number of oxygen functional groups (the slope of the straight line became larger in the low frequency region), and the Warburg impedance increased. On the basis of the study, we also need to further explore the experimental method which can directly observe the energy storage process.